%% file: contribution.tex
\documentclass[sigconf]{acmart}

\AtBeginDocument{%
  \providecommand\BibTeX{{%
    \normalfont B\kern-0.5em{\scshape i\kern-0.25em b}\kern-0.8em\TeX}}}

\setcopyright{acmcopyright}
\copyrightyear{2022}
\acmYear{2022}
\acmDOI{XXXXXXX.XXXXXXX}

\acmConference[CHI '22]{the InContext workshop at the CHI Conference on Human
Factors in Computing Systems}{May 01,
  2022}{New Orleans, LA, USA}
\acmPrice{15.00}
\acmISBN{978-1-4503-XXXX-X/18/06}

\usepackage{pdfcomment}

\begin{document}

\title{Social Practice Cards: Research Material to Study Social Contexts as
Interwoven Practice Constellations}


\author{Alarith Uhde}
\email{alarith.uhde@uni-siegen.de}
\orcid{0000-0003-3877-5453}
\affiliation{%
\institution{University of Siegen}
\streetaddress{Kohlbettstraße 15}
\city{Siegen}
\country{Germany}
\postcode{57072}
}

\author{Mena Mesenhöller}
\email{mena.mesenhoeller@gmx.de}
\affiliation{%
  \institution{Heidelberg University}
  \streetaddress{Grabengasse 1}
  \city{Heidelberg}
  \country{Germany}
  \postcode{69117}
}

\author{Marc Hassenzahl}
\email{marc.hassenzahl@uni-siegen.de}
\orcid{0000-0001-9798-1762}
\affiliation{%
\institution{University of Siegen}
\streetaddress{Kohlbettstraße 15}
\city{Siegen}
\country{Germany}
\postcode{57072}
}

%\renewcommand{\shortauthors}{Uhde and Hassenzahl}

\begin{abstract}

  Studying how social contexts shape technology interactions and how we
  experience them is hard. One challenge is that social contexts are very
  dynamic and shaped by the situated practices of everyone involved. As
  a result, the same human-technology interaction can be experienced quite
  differently depending on what other people around us do. As a first step to
  study interpersonal and interpractice dynamics, we collected a broad range of
  visual representations of practices, such as ``riding a bike'' or ``skipping
  the rope''. This material can be used to further explore how different,
  co-located practices relate to each other.

\end{abstract}






\begin{teaserfigure}
  \pdftooltip{\includegraphics[height=2.6cm]{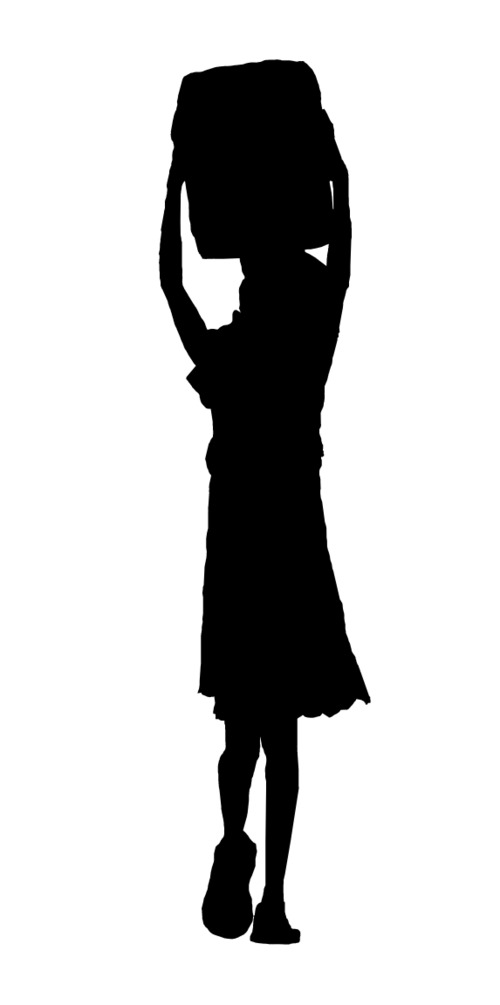}}{Five
  silhouette images of people doing different things. A woman carrying a heavy
  package on her head}
  \pdftooltip{\includegraphics[height=2.6cm]{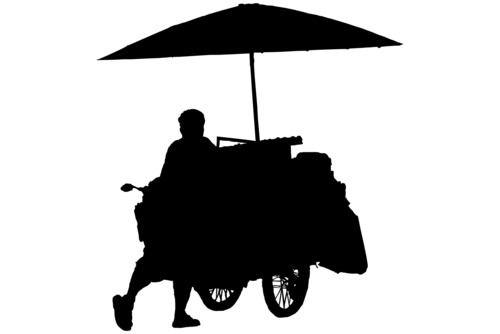}}{A
  man pushing a mobile food stand}
  \pdftooltip{\includegraphics[height=2.6cm]{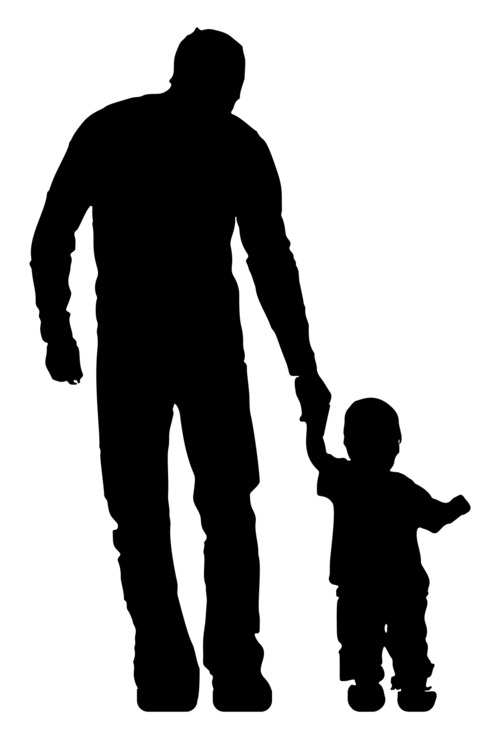}}{A
  father and his young child walking}
  \pdftooltip{\includegraphics[height=2.6cm]{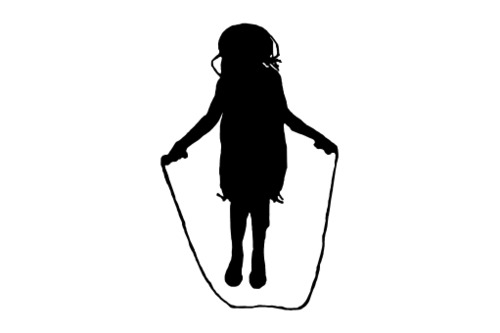}}{A
  girl skipping a rope by herself}
  \pdftooltip{\includegraphics[height=2.6cm]{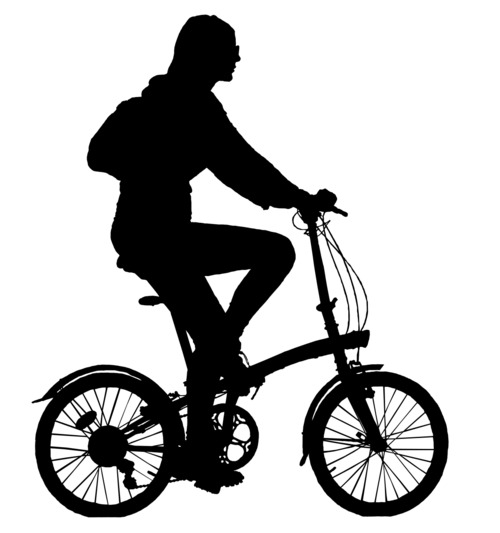}}{A
  woman on a bicycle}
  \pdftooltip{\includegraphics[height=2.6cm]{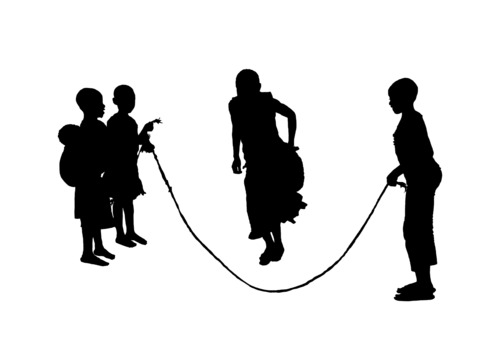}}{A
  group of children skipping rope together}
  \caption{Sample Practices from the Social Practice Cards}
  \Description{Five silhouette images of people doing different things. From
  left to right: A woman carrying a heavy package on her head, a man pushing
  a mobile food stand, a father and his young child walking, a girl skipping a rope
  by herself, a woman on a bicycle, a group of children skipping rope together.}
  \label{fig:teaser}
\end{teaserfigure}

\maketitle

\section{Introduction}

``Social context'' immensely impacts people's experience of interactions with
technology. Sometimes a particular social context is a necessary prerequisite
for particular human-technology interactions, for example, when taking
a souvenir photo of friends. In other cases, social context brings new
experiential qualities to an interaction, as when people play music to others.
While important, Human-Computer Interaction (HCI) is far from a consensus about
what ``social context'' is, beyond the rather vague idea that it is about
situations where people share some space~\citep{Dourish2004, Uhde2021b}. Neither
do we have a comprehensive taxonomy of social situations, nor do researchers
agree upon a set of crucial attributes to describe them. As a consequence, there
are only few tools to systematically study how social context shapes
human-technology interactions and people's experiences.

This work builds upon a conceptualization of social context based on Social
Practice Theory (SPT; e.g.,~\citep{Kuutti2014, Reckwitz2002, Shove2012}), as
intertwined and intersecting constellations of co-located practices. When people
come together, they co-perform several practices (e.g., standing, smiling,
pulling a face, taking a photo). Following SPT, these collective practices and
resulting interactions between them \emph{are} the social context, which
dynamically shapes the single practice performances and the likelihood of new
practices to emerge or not. For example, in most situations, smiling and taking
a photo go together and may even be interdependent. In contrast, other practices
are in conflict with each other (such as ``writing an exam'' and ``loud
chatting''). Finally, some practices can seem completely ``out of context'' for
each other with unclear experiential consequences, because they are rarely
co-performed in the wild (e.g., rope skipping while attending a cocktail party).
% explore as-if scenarios % search for incompatibilities and design for them

In this sense, social contexts can be characterized based on the practices they
allow for or even demand, and the practices they exclude. For example, Erving
Goffman~\citep{Goffman1966}, who studied ``social gatherings'' with a similar
notion, observed that ``a funeral'' can be characterized by the specific set of
required and accepted practices (e.g., mourning, talking with a low voice),
while it also excludes several other practices (e.g., dancing). Of course,
included and excluded practices vary across cultures and over time, and social
practice theory accounts for such differences through local contigencies and
a ``history'' of practices and their associated meanings~\citep{Shove2012}.
Although this general observation seems to resonate with people's subjective
experiences, the exact processes that determine which practices go well together
and which do not remains too vague to be used effectively in design, although
some of these relationships are relatively obvious: The noise produced by
playing loud music wakes people up who are trying to sleep, so the ``playing
music'' practice makes successful ``sleeping'' practices unlikely. In other
cases, culture or traditions, local values, and people's subjective experiences
are important. One example for this is an assumed conflict between taking photos
and behaving naturally that has led some dance clubs to physically cover the
cameras of their guests' smartphones (e.g., with a sticker) to prevent people
from taking photos and to allow for a less disturbed atmosphere.

Based on the practice approach, we believe that a key to a better understanding
of social contexts and how they are interwoven with human-technology
interactions and user experiences lies in a thorough and systematic
understanding of such facilitating and interfering interpractice relationships.
However, to explore them requires the comparison of a substantial number of
social practices combined with each other in different constellations to
establish a particular social context.

We think that a systematic understanding of these relationships between
co-located practices would help designers anticipate design requirements beyond
the current, often decontextualized approach to developing human-technology
interactions. It could also reveal local (in-)compatibilities as opportunities
to reshape practice relationships through design.

To make this more approachable, we collected an initial heterogenous set of 203
practices and represent them as research material. We will use this to study
constellations of practices in future user studies by assembling imagined
contexts, and then gathering as well as categorizing emerging interpractice
relationships. Our longer-term goal is to better understand and model how and
why some practices harmonize with each other, while others do not, and to make
this knowledge more accessible to designers. In this paper, we present the
development process of the ``Social Practice Cards'', the current state of the
set, and our plans to further use and extend it in the future.

\section{Design Considerations}

We began to work on the ``social practice cards'' out of a need for easily
accessible and flexible material to represent social contexts as constellations
of practices. The general idea is to collect representations of human
activities, broadly but also with a focus on human-technology interactions
involved.

We had several requirements for the material:

\begin{itemize}
  \item{Easy to interpret, simple representations}
  \item{Diversity of practices}
  \item{Aesthetic consistency}
  \item{Minimal distraction by unrelated details}
  \item{Easy extensibility}
\end{itemize}

Consequently, we chose a simple visual format that can be used to represent
a wide range of different practices while providing some aesthetic consistency
and authenticity. Photo-based material seemed useful, but it usually includes
several unnecessary or possibly distracting details. Specifically, we wanted to
set the focus on the performed practice more than the represented person. Thus,
simplified, high contrast silhouette images seemed appropriate (see
Figure~\ref{fig:teaser}).

Another design goal was extensibility. Other researchers and designers may have
different needs and should be able to easily study different practices without
having to deal with complicated graphics manipulation or expensive software. The
silhouette format is already established, so there is an abundance of
instructional guides available to create them with several (also free) software
packages. In addition, there is a wide range of silhouette images available
online, many with permissive licenses.

We included a broad range of practices in the initial set, which represent
contemporary practices performed by people all around the world. We hope that
this variability covers a wide range of social contexts, and proves to be
stimulating in future research. That said, we are aware that such a collection
of activities can never capture all possible practices and their nuances.

\begin{figure}[h]
\centering
  \pdftooltip{\includegraphics[width=0.27\linewidth]{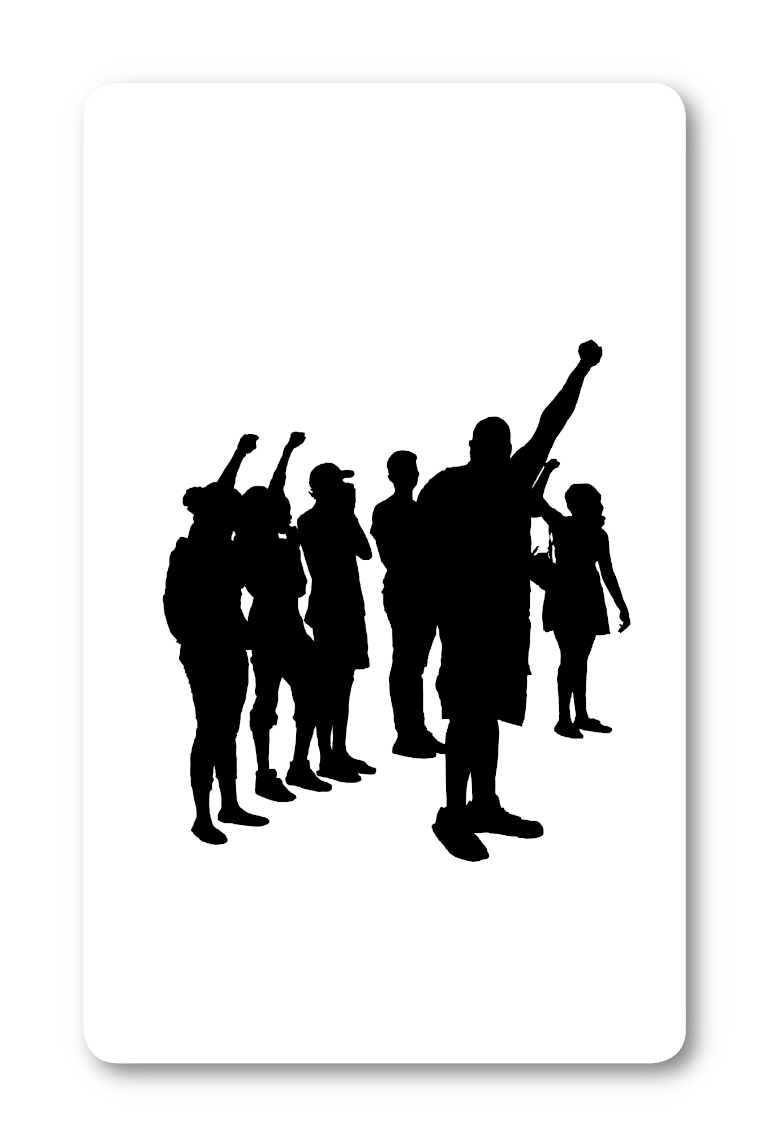}}{The
  silhouette of a crowd of people protesting}
  \pdftooltip{\includegraphics[width=0.27\linewidth]{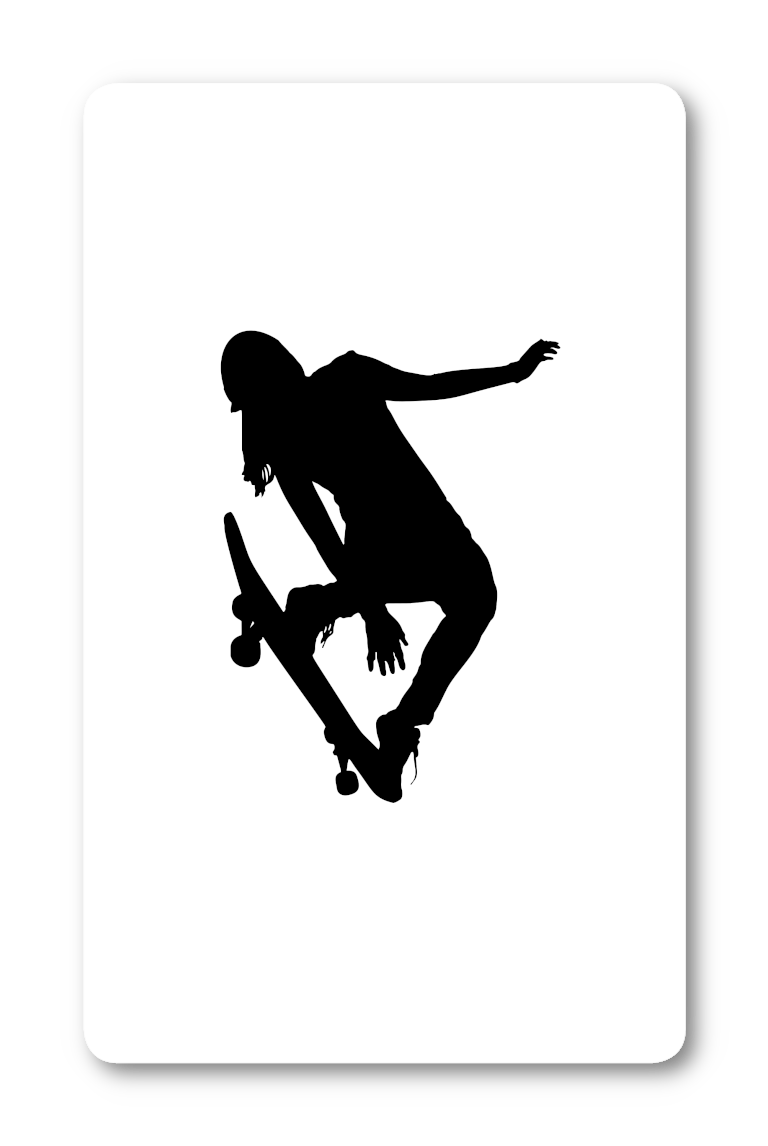}}{A
  young woman performing a trick on a skateboard}
  \pdftooltip{\includegraphics[width=0.27\linewidth]{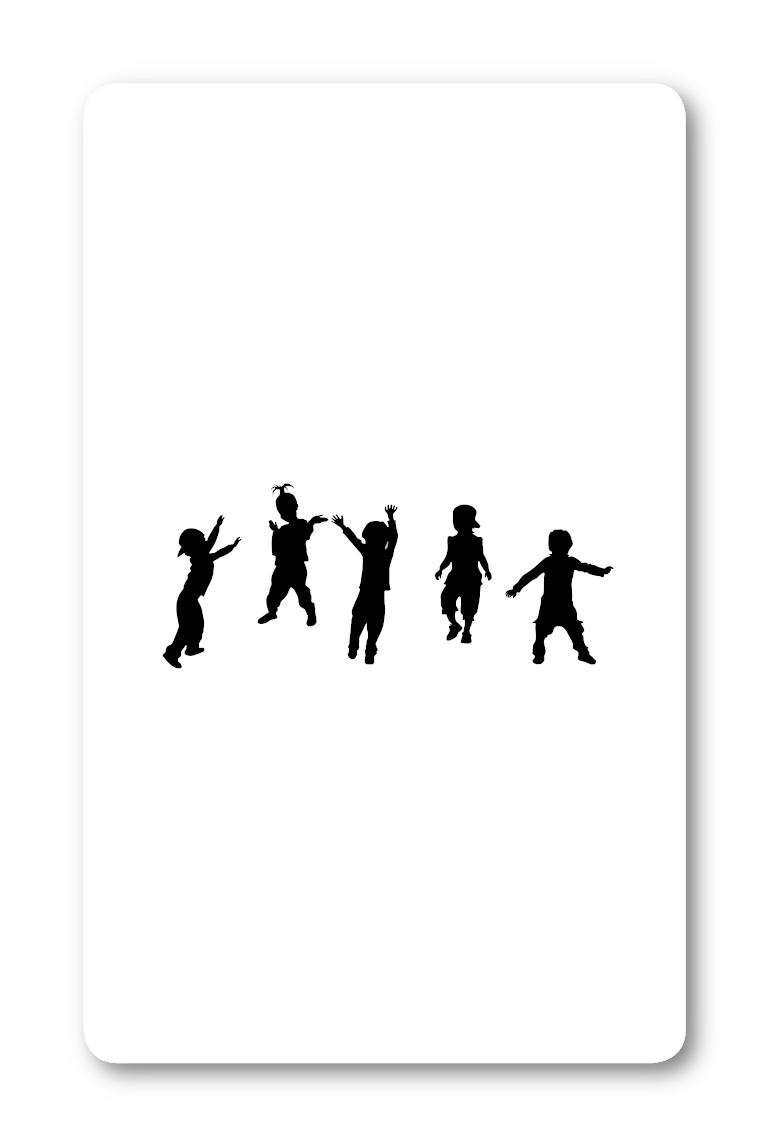}}{A
  group of children jumping and dancing}
\caption{Example cards of three practices.}
\Description{The cards depict a crowd of people protesting, a young woman
  performing a trick
  on a skateboard, and a group of children jumping and dancing.}
  \label{fig:cards}
\end{figure}

\section{Material Collection}

We collected pictures of practices from various online sources. Three people
independently searched for pictures that could either be normal photos or
already transformed into a silhouette format. We had some a priori criteria for
inclusion, but added a few more throughout our search process:

\begin{itemize}
  \item{The full person and practice should be visible (e.g., including their feet)}
  \item{The picture is available under a free license that requires no
    attribution}
  \item{The source picture set includes people with a mix of gender, age, and ethnicity, and
    people with different physical conditions}
  \item{The depicted scene is relatively contemporary and realistic}
\end{itemize}

For some activities, depending on the orientation of the depicted person
relative to the photographer, the activities were not clearly discernible after
their transformation to silhouettes. For example, if the person read a book that
she held between herself and the camera, it disappeared in the silhouette. In
these cases, we searched for an alternative picture of the practice in
a different orientation and included it, if available.

The license restriction was added because it allows us and others to freely
share the set. However, we had to exclude many images that would otherwise fit
well. We hope that we can extend the set in the future with more free images,
but a workaround for user researchers is to include their own, non-free images
in their specific work, without publishing them online.

Finally, some images we found depicted people in historic or otherwise fictional
contexts (e.g., a witch on a broom), in which case we did not include them.
Should such practices be relevant for future researches, they can easily add
them.

\section{Current Status}

The set currently contains a total of 203 silhouette images~\citep{Uhde2022}. We
have organized them according to broad categories as listed in
Table~\ref{tab:categories}. 135 silhouettes represent practices performed alone,
67 silhouettes represent practices performed together with one or more other
people, and one is a picture of a robot without a performing human. When
performing alone, the person is lying in six silhouettes, sitting in 20,
standing in 45, and moving in 65. The categorization was more difficult in the
practices performed by more than one person, and we quickly ended up with many
subcategories. Thus, we decided to only distinguish between spatially dynamic
and static practices, and ``mixed'' ones (i.e., those in which some people are
relatively static while others are more dynamic). We transformed 93 images from
the original source. Most of these transformations were from photo to silhouette
format, in a few cases we have split up a silhouette image with several
seemingly unrelated people into separate practices.

\input{figures/categories.tex}

\section{Further Research}

In our own research, we are particularly interested in further exploring
different types of relationships that exist between practices (see
e.g.,~\citep{Uhde2021b}). We already worked on the positive and negative
interplay between practices based on sound. For example, phone calls seem to
disturb some co-located practices such as reading, but they can also enable new
practices, such as overhearing the call~\citep{Norman2014} and possibly
facilitate entailing conversations. However, sound is only one specific
dimension and others are still relatively unexplored.

Previous research in HCI, especially on social acceptability, has considered
different context ``categories'' (e.g.~\citep{Rico2010}) as an ad hoc measure to
study the experiential differences between social contexts. For example, common
context categories include a sidewalk, a bus, a pub, or a workplace. Of course,
this list of categories was never assembled with the intention to represent all
possible contexts, but rather to explore a broad set of contexts that are still
manageable to cover in typical user studies. Our approach is somewhat different,
because we differentiate contexts based on specific practices of other people
rather than location categories. While this adds some complexity, we believe it
to be fruitful for future work, because it helps us actually understand and
describe ``why'' for example a workplace might shape user experiences in
a different way than a sidewalk. This allows us to study the similarity and
differences between contexts. For example, we would expect a substantial but not
complete overlap between ``bar'' and ``restaurant'' practices, and a lower
overlap between ``bar'' and ``kindergarten'' practices. Assumingly, this overlap
would go along with more similar user experiences of a technology interaction in
these contexts. In addition, we could also better study hybrid settings and how
they relate to their origins (e.g., a board restaurant as a mix between
a restaurant and a train). Compared with the categorical approach, this
practice-based perspective can help us better interpret previous results and
their validity in new contexts. Thus, one of our central goals with the
practice-based approach is to better describe and study social contexts, so we
can make use of them as a resource for design, rather than a complication in
need to be managed.

To study the structure of practice relationships, we currently consider two
methodological approaches for which we plan to use the social practice cards.

First, we plan a card sorting study to understand which practices people believe
to go well together and which not. Card sorting is an efficient way to collect
subjective one-to-one relationships between different elements (here practices).
We plan to use card sorting to identify more specific clusters of practices
which give us a broad overview of the practices people think go well together
and which do not.

Second, we plan to use example practices from the card sorting clusters for
a follow-up repertory grid study~\citep{Jankowicz2005, Kelly1955}. In
a repertory grid study, participants are typically presented with sets of three
elements drawn from a larger pool, in our case visual practice representations.
Their task is to group two of these elements together that they see as similar
in some way, but different from the third, and to describe the difference in
their own terms as a bipolar attribure pair (e.g., round -- cornered). An
advantage of repertory grid is that participants create their own ``scales'', so
their judgments directly relate to their actual experiences. We hope to derive
commonalities between different people's experiences from this, that can inform
further research and design.

In addition, we plan to make our findings available together with the social
practice cards and invite other researchers to share their findings with us.
Over time, this could produce a rich, annotated set of practices that can be
useful in broader research areas.

\section{Practical Uses}

% Konkrete Beispiele
% Den Prozess durchlaufend

In addition to advancing HCI research by systematically exploring interpractice
relationships to model facilitating and interfering relationships and their
impact on user experiences, the social practice cards can be integrated with
a broad range of existing methods in user experience design. We outline a few
examples here that are meant to cover a wider variety of different approaches.

First, in quantitative research approaches, designers could describe their
technology concept (e.g., a technologically enhanced skateboard or scooter) in
an early stage of the design process with a textual vignette format during
a questionnaire study. The silhouettes can then be used in a next step to
imaginarily place this technology in different social situations (represented by
visual cues) and collect data about people's anticipated experiences. This
approach is similar to existing work on social acceptability, but may provide
more nuanced insights because of the more fine-grained context descriptions.

Second, it can also be used in qualitative research to more deeply investigate
how people imagine their subjective user experiences around other people
performing certain practices. Following the skateboard example, this can be
helpful to study the ``why'' behind (in-)compatibilities and explore ways to
work around them. To facilitate such qualitative use cases, we have provided
printable templates in game card format (see Figure~\ref{fig:cards}). In early
pilot tests we found them to facilitate immersion, to invite reassembling new
practice combinations, and to help participants reflect on the practice
relationships.

Third, researchers can focus on specific target contexts and collect typical
local practices in a pilot field study. The social practice cards can be used to
create a context portfolio to guide the further design process (we think of this
as similar to personas, but representing practices in context instead of
people). This can serve as a more thorough representation of the target context
and can support reflection throughout the design process. For example, the
development of social contexts over time based on disappearing and new practices
can be represented explicitly through practice representations and directly
inform further design. As an example, e-scooters are such a relatively new
practice in sidewalk or park contexts in several cities that have recently
started to shape how people interact with each other and their environment in
these contexts.

\section{Conclusion}

With this workshop contribution, we present the ``Social Practice Cards'', a set
of research material to study social contexts based on the relationships between
multiple, co-located practices. We framed social context as a constellation of
(in-)compatible practices, whose relationships shape people's situated
experiences. The presented material helps to flexibly study such relationships,
and how new technology interactions fit in with existing social contexts. It is
based on a simple but versatile visual format, freely available, and easily
extensible. The set currently includes a diverse collection of 203 practices
that can be used in a wide range of qualitative and quantitative research.

\begin{acks}

This project is funded by the \grantsponsor{501100001659}{Deutsche
  Forschungsgemeinschaft (DFG, German Research
  Foundation)}{https://doi.org/10.13039/501100001659} – Grant
  No.~\grantnum{425827565}{425827565} and is part
  of~\grantnum{427133456}{Priority Program SPP2199 Scalable Interaction
  Paradigms for Pervasive Computing Environments}. We would like to thank Kieu
  Tran for his support during the image collection.

\end{acks}

\bibliographystyle{ACM-Reference-Format}
\bibliography{bibliography.bib}

\end{document}

%% file: figures/categories.tex
\begin{table}[h]
  \caption{Categorization of Silhouettes according to some formal criteria. At
  the bottom of the table we have included summaries for each criterion.}
  \label{tab:categories}
  \begin{tabular}{lllc}
    \toprule
    Performers & Posture  & Group  & Silhouettes \\
    \midrule
    alone      & lying    & child  &   2 \\
    alone      & lying    & female &   2 \\
    alone      & lying    & male   &   2 \\
    alone      & sitting  & child  &   2 \\
    alone      & sitting  & female &  11 \\
    alone      & sitting  & male   &   7 \\
    alone      & standing & child  &   4 \\
    alone      & standing & female &  12 \\
    alone      & standing & male   &  27 \\
    alone      & moving   & child  &   9 \\
    alone      & moving   & female &  23 \\
    alone      & moving   & male   &  33 \\
    alone      & other    & female &   1 \\
    together   & dynamic  & child  &   7 \\
    together   & dynamic  & female &   5 \\
    together   & dynamic  & mixed  &  29 \\
    together   & static   & child  &   3 \\
    together   & static   & female &   2 \\
    together   & static   & male   &   7 \\
    together   & static   & mixed  &  13 \\
    together   & mixed    & mixed  &   1 \\
    technology &          &        &   1 \\[0.1in]
    child      &          &        &  27 \\
    female     &          &        &  56 \\
    male       &          &        &  76 \\
    mixed      &          &        &  43 \\[0.1in]
    alone      &          &        & 135 \\
    together   &          &        &  67 \\[0.1in]
    lying      &          &        &   6 \\
    sitting    &          &        &  20 \\
    standing   &          &        &  43 \\
    moving     &          &        &  65 \\
    dynamic    &          &        &  41 \\
    static     &          &        &  25 \\
  \bottomrule
\end{tabular}
\end{table}